\documentclass[12pt]{iopart}
\usepackage{amsfonts}
\usepackage{lscape}

\begin{document}
\input{epsf.tex}
\begin{center}\Large The XX--model with boundaries.\\ Part II: 
Finite size scaling and partition functions
\\[1cm]
\normalsize
Ulrich Bilstein\footnote{E-mail address: \tt bilstein@theoa1.physik.uni-bonn.de}\\[0.5cm]
 Universit\"{a}t Bonn \\
                    Physikalisches Institut, Nu\ss allee 12,
                    D-53115 Bonn, Germany\\[1.3cm]
\end{center} 
{\bf \small Abstract.}
\small
We compute the continuum limit of the spectra 
 for the XX--model with arbitrary complex
boundary fields. 
In the case of hermitian boundary terms one obtains the partition functions of the free 
compactified boson field on a cylinder with  
Neumann-Neumann, Dirichlet-Neumann or Dirichlet-Dirichlet 
boundary conditions.  
This applies also for certain non-hermitian boundaries.
For special cases we also compute the free surface energy.
For certain non-hermitian boundary terms the results are more complex.
Here one obtains logarithmic corrections to the free surface energy.
The asymmetric version of the XX--model with boundaries (this includes the Dzyaloshinsky-Moriya
interaction)
is also discussed.\\[0.2cm]
\normalsize
%\pacs{007, 08.15}
\section{Introduction}
In the first paper of this series \cite{paper1} we  have shown 
how to  diagonalize the  Hamiltonian of the XX--model with boundaries
\begin{equation}
\fl
H=\frac{1}{2}\sum_{j=1}^{L-1}
\left[\sigma_j^+\sigma_{j+1}^-+\sigma_j^-\sigma_{j+1}^+\right]
+\frac{1}{\sqrt{8}}\left[\alpha_-\sigma_1^-+\alpha_+\sigma_1^+
+\alpha_z\sigma_1^z+\beta_+\sigma_L^++\beta_-\sigma_L^-+\beta_z\sigma_L^z\right]
\label{HXX}
\end{equation}
by introducing an auxiliary Hamiltonian 
\begin{eqnarray}
\fl
H_{\rm long}=\frac{1}{2}\sum_{j=1}^{L-1}
\left[\sigma_j^+\sigma_{j+1}^-+\sigma_j^-\sigma_{j+1}^+\right] \nonumber \\
+\frac{1}{\sqrt{8}}\left[\alpha_-\sigma_0^x\sigma_1^-+\alpha_+\sigma_0^x\sigma_1^+
+\alpha_z\sigma_1^z+\beta_+\sigma_L^+\sigma_{L+1}^x+\beta_-\sigma_L^-\sigma_{L+1}^x
+\beta_z\sigma_L^z\right]
\label{Hlong}
\end{eqnarray}
which in turn may be diagonalized
in terms of free fermions. The parameters $\alpha_{\pm},\beta_{\pm},\alpha_z$ and $\beta_z$
are arbitrary complex numbers.
Note that $H_{\rm long}$ commutes with $\sigma_0^x$ and $\sigma_{L+1}^x$.
Hence the spectrum of $H_{\rm long}$ decomposes into four sectors 
$(+,+),(+,-),(-,-),(-,+)$ corresponding to the eigenvalues $\pm 1$ of $\sigma_0^x$
and $\sigma_{L+1}^x$. 
The spectrum of  $H$ is obtained by projecting onto the $(+,+)$--sector.

While in \cite{paper1} our focus was on the diagonalization
of the finite chain, here
 we are going to examine the asymptotic behaviour of 
 the energy levels for large values of $L$.
In the case of conformal invariance 
this information is usually encoded in the partition function,
 which we define by
\begin{equation}
{\cal Z}= \lim_{L\to \infty}\tr z^{ \frac{L}{\xi}(H-e_{\infty}L-f_{\infty})}
\label{partf}
\end{equation}
where $z<1$ and $e_{\infty}$ and $f_{\infty}$ denote the free bulk and the free
surface energy respectively. 
$\xi$ may be considered as a normalization constant here. We will also consider the partition
function for $H_{\rm long}$.

In the case of periodic boundary conditions the continuum limit of the XXZ--chain can
be described by a free boson field $\Phi$ (see e.g. \cite{ARev}), with action
\begin{equation}
S=\frac{1}{2}\int dx_1 dx_2 \left[ \left(\partial_1 \Phi\right)^2+
			     \left( \partial_2 \Phi\right)^2 \right]
\end{equation}
being compactified on a circle of radius $r$, i.e. we identify 
$\Phi$ with $\Phi+2\pi r$ (we follow the notation of \cite{Sal}).
The radius of compactification $r$ is related to the value of the
anisotropy in the XXZ--model. At the free fermion-point the anisotropy 
is zero and we have $r=1/\sqrt{4\pi}$.

Consider the boson field  on a cylinder of length $L$ and
circumference $N$ imposing periodic boundary conditions in the direction
of the cylinders length.
Quantizing the field with time in $N$-direction 
  yields the partition
function \cite{Sal}
\begin{equation}
\label{bosonp}
{\cal Z}_{\rm periodic}=\frac{1}{\eta(q)^2}\sum_{m,n \in \mathbb{Z}} 
	q^{\frac{1}{2}\left( \frac{m^2}{2\pi r^2} + 2\pi r^2 n^2\right)}
\end{equation}
where $q=\rme^{-2\pi N/L}$ and 
\begin{equation}
\eta(q)=q^{\frac{1}{24}}\prod_{n=1}^{\infty}(1-q^n).
\end{equation}
This expression coincides with the partition function
 for the periodic XXZ-chain, which has been studied 
extensively in \cite{AlcBaaGriRit2,AlcBarBat,AlcBarBat2}.
The partition function for $H_{\rm long}$ in the periodic case 
is just $4 {\cal Z}_{\rm periodic}$.

Suppose now the case of open boundaries at both ends of the cylinder.
 There are  two different 
 boundary conditions preserving conformal invariance,
 i.e. the Dirichlet and the von Neumann (see e.g. \cite{Sal,OshAff} ).
This yields three different partition functions corresponding 
to the various possible combinations of the two boundary conditions.
The respective partition functions can be found in \cite{Sal,OshAff}.
Imposing Dirichlet boundary conditions at both ends of the
cylinder results in
\begin{equation}
\label{bosondd}
{\cal Z}_{\rm DD}(\Delta) =\frac{1}{\eta(q)}\sum_{n \in \mathbb{Z}}
	q^{\frac{1}{2} (2 \sqrt{\pi} r n+\Delta)^2	}
\end{equation}
where $q=\rme^{-\pi N/L}$ and $\Delta=(\Phi_0-\Phi_L)/\sqrt{\pi}$.
By $\Phi_0$ and $\Phi_L$ we denote the values of
the boson field at the boundaries.
This type of partition function 
 has also been obtained for the open XXZ-chain 
with diagonal, hermitian boundary fields \cite{AlcBaaGriRit1}.

The Neumann-Neumann partition function 
is given by
\begin{equation}
\label{bosonnn}
{\cal Z}_{\rm NN}(\Delta) =\frac{1}{\eta(q)}\sum_{m \in \mathbb{Z}}
        q^{2(m/(2\sqrt{\pi}r) +\Delta/2)^2 }  
\end{equation}
where $\Delta=(\tilde{\Phi}_0-\tilde{\Phi}_L)/\sqrt{\pi}$ and $\tilde{\Phi}_0,\tilde{\Phi}_L$ denote the values
of the dual field $\tilde{\Phi}$ at the boundaries \cite{Sal,OshAff}.
The Dirichlet-Neumann boundary condition yields \cite{Sal,OshAff}
\begin{equation}
\label{bosondn}
{\cal Z}_{\rm DN} =\frac{1}{2\eta(q)}\sum_{k \in \mathbb{Z}}
        q^{\frac{1}{4} (k-1/2)^2 }  .
\end{equation}
In this case no free parameter appears. 

Which boundary conditions of the Hamiltonian $H$ lead to the partition functions
\eref{bosonnn} and \eref{bosondn} is yet not known.
We will close this gap. 
We are going to see that, as long as the Hamiltonian $H$ is  hermitian, the partition
function of the chain corresponds to one of the three  boson partition
functions for open boundaries. However, this may also be the case for certain non-hermitian
boundary terms.
See section 2 for a survey over the boundary conditions which 
we will consider.

In \cite{Aff2}  it has been argued that  non-diagonal  boundary terms
correspond to von Neumann boundary conditions
for the boson field. Our results will show that this assumption is correct
in the hermitian case, but not necessarily for non-hermitian boundaries.

In this paper we also study the asymmetric XX--chain with boundaries (see equation
\eref{Ha}). This includes the quantum chain with Dzyaloshinsky-Moriya interactions,
which was already studied in \cite{AlcWre} for periodic boundary conditions. 
The partition function for real values of $p$ and $q$ in \eref{Ha} and periodic boundary conditions was 
obtained in \cite{NohKim}.
The special case $p=1$ and $q=0$ with boundaries has already been studied in \cite{paper0}.

This paper is organized as follows: We will start with a summary of our results for
the symmetric chain with hermitian and non-hermitian boundaries in section 2. The
asymmetric XX-chain will be discussed in section 3. We will give our conclusions in section 4.
The appendix is dedicated to the computation of the fermion energies which yield 
the energy gaps of the chain.

\section{Summary of results}
The structure of our results
is most simply encoded in terms of new parameters $F,C,G,K,J$, where
\begin{eqnarray}
\label{newparam}
\fl
F=4 \alpha_-\alpha_+\beta_+\beta_- \nonumber \\
\fl
C=\alpha_-^2\beta_+^2+\alpha_+^2\beta_-^2  \nonumber \\
\fl
G =2\alpha_-\alpha_+(1+2\beta_z^2)+2\beta_-\beta_+(1+2\alpha_z^2)
\nonumber \\
\fl
K =2(\alpha_-\alpha_++\beta_-\beta_+)+(1+2\alpha_z^2)(1+2\beta_z^2)\nonumber \\
\fl
J =8\alpha_z\beta_z-4\alpha_z^2\beta_z^2+2(\alpha_-\alpha_++\beta_-\beta_+) +2\alpha_z^2+2\beta_z^2
-1 .
\end{eqnarray}
The six different situations we considered are given in 
\tref{structure}. 
This table also indicates which kind of partition function we obtained for $H$.
For hermitian boundaries the cases in this table cover the whole parameter 
space. This is not the case for non-hermitian boundary terms.

\begin{table}
\caption{Conditions on the parameters $F,C,G,K,J$ being considered in this paper.
 An asterisk indicates that the corresponding value may be arbitrary.
The partition functions ${\cal Z}_{DD},{\cal Z}_{NN}$ and ${\cal Z}_{DN}$ 
are defined in \eref{bosondd},\eref{bosonnn} and \eref{bosondn}.
For the last two cases we obtained anomalous behaviour of the energies (see section 2.5).
We use h. and n-h. as a shortcut for hermitian and non-hermitian.
}
\begin{indented}
\item[]
\begin{tabular}{@{}lccccrr}
\br
F  & C & G & K & J &  &  \\
\mr
$\neq 0$ & * & * & * & * &  ${\cal Z}_{NN}$ & h. and n-h.  \\
0 &  0 & $\neq 0$ & * & * & ${\cal Z}_{DN}$ & h. and n-h. \\
0 &  0 & 0 & $\neq 0$  & *  & ${\cal Z}_{DD}$ & h. and n-h. \\
0 &  0 & 0 & 0 & 0 &  $2\times{\cal Z}_{DD}$  & n-h. \\ \mr
0 &  $\neq0$ & $\neq 0$ & *  & * & anom. & n-h. \\
0 &  0 & 0 & 0 & $\neq 0$ & anom. &  n-h. \\
\br
\end{tabular}
\end{indented}
\label{structure}
\end{table}
For the cases in \tref{structure} we computed analytically the energy gaps in the leading
order as shown in appendix A. The exact ground-state energies for certain cases being
listed in \tref{tabfac} have already been obtained in \cite{paper1}.

However, the information of the energy gaps is already sufficient to obtain the
partition functions for $H_{\rm long}$ upto a factor $z^{-\frac{1}{24}+h_{\rm min}}$, where
$h_{\rm min}$ denotes the lowest highest weight appearing in the representation
of the Virasoro algebra.
This information has to be extracted from the ground-state energies \cite{Car}
\begin{equation}
\label{gstate}
E_0=e_{\infty}L+f_{\infty}-\frac{1}{L}\left(\frac{1}{24}-h_{\rm min}\right)+ \ldots
\end{equation}
The value of the free bulk-energy is already known from the periodic chain, i.e. $e_{\infty}=-\frac{1}{\pi}$ 
\cite{LSM}. The values of the free  surface energies $f_{\infty}$ and of $h_{\rm min}$ have been obtained 
for the cases given in \tref{tabfac} expanding the exact expressions for the finite chain given in \cite{paper1}.

The results confirm the partition functions we will give for $H_{\rm long}$ in the following.
For the cases not being listed in \tref{tabfac} we checked our expressions numerically. 
The partition functions for $H$ have been obtained afterwards by projecting onto the $(+,+)$-sector
(see \cite{paper1} for details on the projection mechanism). We are now going 
to discuss the cases given in \tref{structure} in detail.
\subsection{Neumann-Neumann boundary conditions}
The condition $F\neq 0$ simply implies that all non-diagonal 
boundary terms are present.
 The diagonal terms 
may be arbitrary. It is quite instructive to introduce a new parametrisation
of the boundary parameters $\alpha_{\pm}$ and $\beta_{\pm}$ as follows:
\begin{equation}
\alpha_+=R_{\alpha}\rme^{\rmi\pi\varphi}\quad \alpha_-=R_{\alpha}\rme^{-\rmi
\pi\varphi}\quad
      \beta_+=R_{\beta}\rme^{\rmi\pi(\chi+\varphi)}\quad \beta_-=R_{\beta}\rme^{-\rmi\pi(\chi+\varphi)}
\label{parann}
\end{equation}
where $R_{\alpha},R_{\beta} \in \mathbb{R}^+,\chi,\varphi  \in \mathbb{R},-1<\chi\leq 1$.
The value of $\Delta$ which enters the partition function \eref{bosonnn} is then
given by 
\begin{equation}
\label{deltann}
\Delta=\left\{ \begin{array}{ll} \chi & \mbox{for $L$ odd} \\
			 \chi+1 &  \mbox{for $L$ even} \end{array} \right. .
\end{equation}
\begin{landscape}
\fulltable{ The values of the parameters $F,C,G,K,J$ for
the cases where the ground-state energy of $H_{\rm long}$ has been obtained in \cite{paper1}. \\
The table shows the free surface energy $f_{\infty}$ and the lowest highest weight $h_{\rm min}$ (cf. \eref{gstate}). The 
expressions for $f_{\infty}$ are also valid for $H$.\\
$\zeta$ denotes a free parameter, which has to be chosen such that $0\leq \mbox{Re}\zeta \leq 2\pi$
for case 12
 and  $-\pi<\mbox{Re} \zeta<\pi$ otherwise .
 }
\footnotesize
\label{tabfac}
\begin{tabular}{@{}lcccccccc}
\br
&&&&&&& \centre{2}{$h_{\rm min}$} \\
&&&&&&& \crule{2} \\
case & $F$ & $(-1)^L C$ & $G$ & $H$ & $J$ & $f_{\infty}$ & $L$ even & $L$ odd \\
\mr
1 & 2 & 0 & 5 & 4 & 2 & $1/4-3/(2\pi)$ & \multicolumn{2}{c}{$1/32$} \\
2 & 0 & 0 & 2 & 3 & 1 &  $1/4-4/\pi$ & \multicolumn{2}{c}{$1/16$}  \\
3 & 0 & 0 & 1 & 2 & 0 &  $1/2-3/{2\pi}$ & \multicolumn{2}{c}{$1/16$}  \\
4 & 0 & 0 & 4 & 4 & 2 &   $-1/(2\pi)$ & \multicolumn{2}{c}{$1/16$}   \\
5 & 1 & $\frac{1}{2}$ & 3 & 3 & 1 & $1/2-2/\pi$ &  $0$ & $1/8$ \\
6 & 1 & $-\frac{1}{2}$ & 3 & 3 & 1 & $1/2-2/\pi$ &  $1/8$ & $0$  \\
7 & 2 & $1$ & 5 & 4 & 2 & $1/4-3/(2\pi)$ & $0$ & $1/8$ \\
8 & 2 & $-1$ & 5 & 4 & 2 & $1/4-3/(2\pi)$ & $1/8$ & $0$ \\
9 & 4 & $ (-1)^L2 \cos\zeta $ & 8 & 5 & 3  & $-1/\pi$ & \multicolumn{2}{c}{$\zeta^2/(8\pi^2)$}\\
10 & $ 2+2\cos \zeta $ & $1+\cos \zeta $ & $4 +4 \cos \zeta$  &
$3+2\cos \zeta$ & $1+\cos \zeta$ &  $1/2-1/\pi-\cos(\zeta/2)/2$ & $0$ & $1/8$  \\
11 & $4 +4\cos\zeta$ & 0 & $6+6\cos\zeta$ & $4+2\cos\zeta$ &
$2+2\cos\zeta$ &  $1/4-1/(2\pi)-\cos(\zeta/2)/2$ &  \multicolumn{2}{c}{$1/32$} \\
12 & $2-2\cos\zeta$ & $\cos\zeta-1$ & $2-2\cos\zeta$ &
4 & 4 &  $1/2-(\cos(\zeta/4)+\sin(\zeta/4))/2$ & $1/8$ & $0$  \\
13 & $4+4\cos\zeta$ & $-2-2\cos\zeta$ & $4+4\cos\zeta$ &
 $4+2\cos\zeta$ & $4+2\cos\zeta$ & $1/2-1/(2\sqrt{2})-\cos(\zeta/2)/2$ & $1/8$ & $0$ \\
14 &  $8+8\cos\zeta$ & $-2-2\cos\zeta$ & $8+8\cos\zeta$ &
 $6+2\cos\zeta$ & $2\cos\zeta-2$ & $-\cos(\zeta/2)/2$ & $0$ & $1/8$  \\
15 & $8+8\cos\zeta$ & $-4-4\cos\zeta$ & $6+6\cos\zeta$ &
$6+2\cos\zeta$ & $6+2\cos\zeta$ & $-\cos(\zeta/2)/2$ & $1/8$ & $0$    \\
16 & 0 & 0 & 0 &
$2+2\cos\zeta$ & $2+2\cos\zeta$ & $1/2-\cos(\zeta/2)/2$ & $1/8$ & $0$  \\
\br
\end{tabular}
\label{tabfuck}
\endfulltable
\end{landscape}

The partition function for $H_{\rm long}$ is given by 
\begin{equation}
\label{zlongnn}
{\cal Z}_{\rm long}=
2 {\cal Z}_{NN}(\chi)+2{\cal Z}_{NN}(\chi+1)
\end{equation}
 for even and odd values of $L$.
Note that the parametrization \eref{parann} does not work for non-hermitian boundaries.
In this case we may define 
the parameter $\chi$ via
\begin{equation}
\chi=\frac{\rmi}{2\pi} \ln\left( \frac{\alpha_-\beta_+}{\alpha_+\beta_-}\right)
\end{equation}
where the logarithm is taken in a way such that
 $0\leq|\mbox{Re}(\chi)|\leq\frac{1}{2}$
for Re$(\alpha_+\beta_-+\alpha_-\beta_+) > 0$ and $1\geq |\mbox{Re}(\chi)| \geq \frac{1}{2}$
for Re$(\alpha_+\beta_-+\alpha_-\beta_+) < 0$. This rule  is consistent with
the parametrisation \eref{parann},
which implies Re$(\alpha_+\beta_-+\alpha_-\beta_+)=2R_{\alpha}R_{\beta}\cos( \chi\pi)$.
However,  we cannot perform the projection onto the $(+,+)$-sector in general if
the boundaries are non-hermitian. In this case 
our result for the partition function of $H$ is restricted to the cases listed in \tref{tabfac}
which satisfy
 the conditions $\alpha_z=\beta_z=0$ or $\alpha_+=\alpha_-,\beta_+=\beta_-$ \cite{paper1}.
The result for $H_{\rm long}$ is valid in general.
\subsection{Dirichlet-Neumann boundary conditions}
The conditions for the Dirichlet-Neumann case yield
\begin{equation}
\beta_{\pm}=0 \quad \alpha_{\pm}\neq 0  \quad \beta_z\neq \pm\frac{\rmi}{\sqrt{2}}  \quad
\mbox{or}\quad
\beta_{\pm}\neq0 \quad \alpha_{\pm}= 0 \quad \alpha_z\neq \pm\frac{\rmi}{\sqrt{2}} .
\label{dnnonhermparam}
\end{equation}
The value of $\alpha_z$ respectively of  $\beta_z$ may be arbitrary.
The partition function for $H_{\rm long}$ is just ${\cal Z}_{\rm long}=4 {\cal Z}_{DN}$
for $L$ even and odd.
The case of one non-diagonal boundary at the end of a semi-infinite XX--chain
has already been considered in \cite{Gui}.
The author has shown, that the model can be decoupled into
two Ising models
with different boundary conditions. One of them being subject to a free boundary
condition at the end,  the other one  being subject to
the fixed boundary condition.
 The partition functions for the Ising model with
free and mixed boundary conditions are well known \cite{Car,BugSha}.
Taking the product of them also results in \eref{bosondn}.
\subsection{Dirichlet-Dirichlet boundary conditions}
The Dirichlet-Dirichlet partition function is obtained for 
two different types of boundary terms.
The first type is given by 
\begin{equation}
\label{ddnonherm}
\beta_z\neq \pm \frac{\rmi}{\sqrt{2}} \quad \alpha_z\neq \pm \frac{\rmi}{\sqrt{2}}
\quad \beta_-=\alpha_-=0.
\end{equation}
where instead of the last equation we might have also chosen $\beta_+=\alpha_+=0$.
However, for this type of boundaries the non-diagonal boundary terms 
have no influence on the energies even 
on the finite chain.
The second type is given by
\begin{equation}
\label{dd2}
\beta_{\pm}=0 \quad \alpha_{\pm}\neq 0 \quad \beta_z=\pm\frac{\rmi}{\sqrt{2}}
\quad \mbox{or} \quad
\beta_{\pm}\neq 0 \quad \alpha_{\pm}= 0 \quad \alpha_z=\pm\frac{\rmi}{\sqrt{2}}
\end{equation}
where the values of $\alpha_z$ respectively $\beta_z$ may be chosen arbitrarily.
Note that for this second type the boundary terms are non-hermitian.
For both types of boundaries the value of $\Delta$ which enters \eref{bosondd} is given by
\begin{equation}
\label{deltadd}
\Delta=\frac{1}{2\pi}\mbox{arccos}\left( (-1)^{L+1}\frac{J}{K}\right) .
\end{equation}
The partition function for $H_{\rm long}$ is ${\cal Z}_{\rm long}=4 {\cal Z}_{DD}(\Delta)$ 
for even and odd $L$.

\subsection{$F=C=G=K=J=0$}
For the case $F=C=G=K=J=0$ the boundary
parameters are nearly fixed,
i.e.
\begin{equation}
\alpha_z=\pm\frac{\rmi}{\sqrt{2}} \quad \beta_z=\mp\frac{\rmi}{\sqrt{2}}
\label{pa}
\end{equation}
and $\alpha_+=\beta_+=0$ or $\alpha_-=\beta_-=0$.
This case is special.
The partition function we obtained does not fit
into the picture suggested by the analogy to the free boson.
Here an additional zero mode appears. 
The value of $\Delta$ is given by $\Delta=0$ for odd $L$ and by $\Delta=1/2$ for even 
$L$. The partition function for $H_{\rm long}$ is ${\cal Z}_{\rm long}=8{\cal Z}_{DD}(\Delta)$
for $L$ even and $L$ odd.

\subsection{Anomalous behaviour}
\begin{figure}[tbp]
\setlength{\unitlength}{1mm}
\def\setl{ \setlength\epsfxsize{8.0cm}}
\begin{picture}(155,90)
\put(28,80){
        \makebox{
                \setl
                \epsfbox{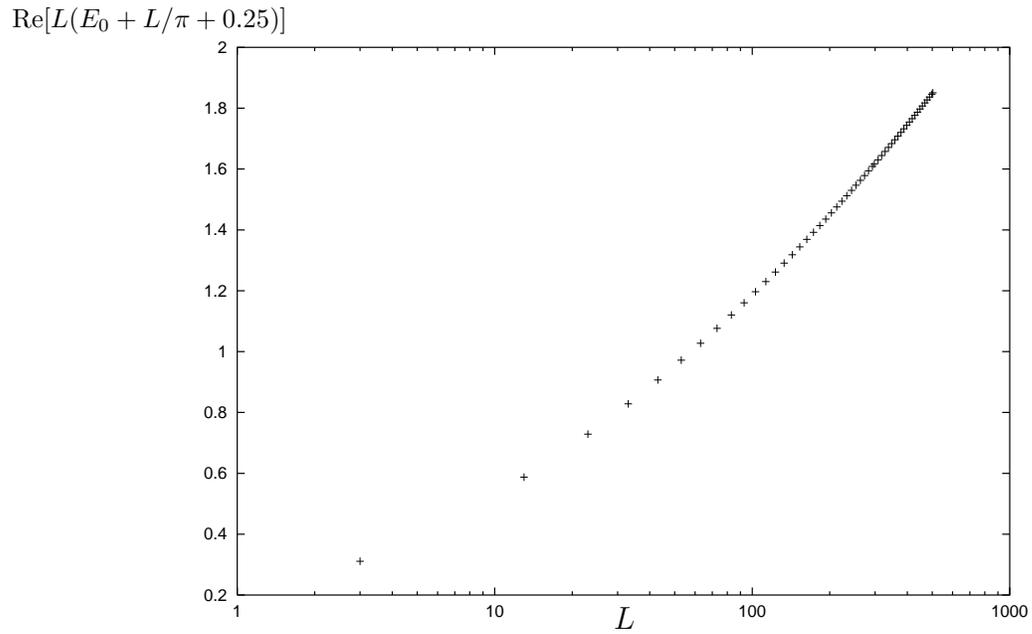}}
        }
\put(10,80){\footnotesize Re$[ L (E_0+L/\pi+0.25)]$}
\put(90,0){$L$}
\end{picture}
\caption{ Real part of the  Casimir amplitude
	$(\alpha_z=1/\sqrt{2},\beta_z=\rmi/\sqrt{2},\alpha_{\pm}=\beta_{\pm}=0)$.}
\label{re}
\end{figure}

\begin{figure}[tbp]
\setlength{\unitlength}{1mm}
\def\setl{ \setlength\epsfxsize{8.0cm}}
\begin{picture}(155,90)
\put(28,80){
        \makebox{
                \setl
                \epsfbox{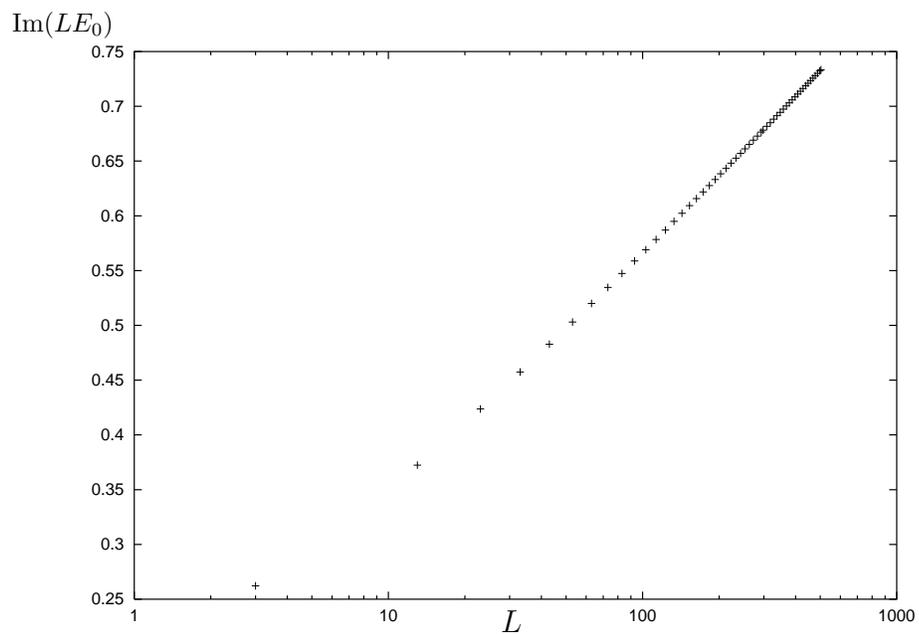}}
        }
\put(25,80){\footnotesize Im$( L E_0)$}
\put(90,0){$L$}
\end{picture}
\caption{Imaginary part of the Casimir amplitude
	 $(\alpha_z=1/\sqrt{2},\beta_z=\rmi/\sqrt{2},\alpha_{\pm}=\beta_{\pm}=0)$.}
\label{im}
\end{figure}

For the last two cases in \tref{structure} we obtained logarithmic 
corrections to the free surface energy.
Note that there is no case in \tref{tabfac} which corresponds 
to the boundaries in question.
However, we computed the ground-state energy $E_0$ of  $H_{\rm long}$ 
numerically for one example of boundaries which lead to the anomalous 
behaviour. Our computations suggest 
\begin{equation}
\label{loge0}
E_0\sim -\frac{L}{\pi} +f_{\infty} +\frac{1}{L}\left(  \rho_1 \ln L + \rho_2 \right)
\end{equation}
where $\rho_1,\rho_2$ are complex numbers. \Fref{re} and \fref{im} show the
real respectively the imaginary part of the Casimir amplitude.
The computation has been done for 
$\alpha_z=1/\sqrt{2},\beta_z=\rmi/\sqrt{2},\alpha_{\pm}=\beta_{\pm}=0$ which corresponds 
to the last case in \tref{structure}.
In both cases the fermionic energies scale as 
\begin{equation}
\label{aslog}
2\Lambda \sim \frac{1}{L} \left[ k\pi \pm \frac{1}{2}\left( \arg \Delta -
 \rmi\ln L \right)\right]  \qquad k=0,1,2,....
\end{equation}
where the value of $\Delta$ is given by
\begin{equation}
\Delta=-\frac{4C}{G} 
\end{equation}
for $F=0,C,G\neq 0$ respectively by
\begin{equation}
\Delta=(-1)^L\frac{2J}{1-4\alpha_z^2\beta_z^2}
\end{equation}
for $F=C=G=K=0,J\neq 0$.
Our result for the last case makes only sense for $(\alpha_z\beta_z)^2\neq 1/4$.
The factor 2 in \eref{aslog} is just a remnant from our notation in \cite{paper1}.
Note  the integer spacing of the energy gaps \eref{aslog}.

\section{The asymmetric XX-chain}
We also considered
the asymmetric XX-chain which is defined by the Hamiltonian 
\begin{eqnarray}
\label{Ha}
\fl
H_{\rm a}=\sum_{j=1}^{L-1}\left[ p \sigma_j^+\sigma_{j+1}^- + q \sigma_j^-\sigma_{j+1}^+ \right]
\nonumber \\
\lo+\frac{1}{\sqrt{8}}\left[
\alpha_-'\sigma_1^-+\alpha_+'\sigma_1^+ +\alpha_z\sigma_1^z  
+\beta_-'\sigma_L^-+\beta_+'\sigma_L^+ +\beta_z \sigma_L^z\right] .
\end{eqnarray}
Without loss of generality we restrict ourselves to $\sqrt{pq}=\frac{1}{2}$.
Under this condition $H_{\rm a}$ can be mapped on the symmetric chain \eref{HXX}
with boundary parameters
\begin{equation}
\label{boundaryDM}
\alpha_-=Q^{\frac{1-L}{2}}\alpha_-' \quad \alpha_+=Q^{\frac{L-1}{2}}\alpha_+'
\quad \beta_-=Q^{\frac{L-1}{2}}\beta_-' \quad \beta_+=Q^{\frac{1-L}{2}}\beta_+'
\end{equation}
by a similarity transformation,
where $Q=\sqrt{\frac{q}{p}}$.
The diagonal terms remain unchanged.
We will separate two cases in the following. First we will consider
  hermitian bulk terms, i.e. 
\begin{equation}
\label{kk}
p=\rme^{\rmi\pi \kappa}/2, q=\rme^{-\rmi\pi \kappa}/2. 
\end{equation}
This corresponds to Dzyaloshinsky-Moriya type interactions in the bulk.
Thereafter we will turn to non-hermitian bulk terms with $p,q$ in $\mathbb{R}$.
\subsection{Dzyaloshinsky-Moriya interactions}
We restricted  ourselves to  hermitian boundaries.
In this case the value of $\kappa$  has only  an  effect onto  the  spectra if
non-diagonal boundary  terms  are  present  at  both ends of the   chain.
We  introduce the parametrization
\begin{equation}
\alpha_+'=R_{\alpha}'\rme^{\rmi\pi\varphi'}
\quad \alpha_-'=R_{\alpha}'\rme^{-\rmi\pi \varphi'}\quad
      \beta_+'=R_{\beta}'\rme^{\rmi\pi(\chi'+\varphi')}
\quad \beta_-'=R_{\beta}'\rme^{-\rmi\pi(\chi'+\varphi')}.
\label{paradm}
\end{equation}
According to \eref{boundaryDM} and \eref{parann} the mapping onto symmetric bulk terms yields
\begin{equation}
\label{prim}
R_{\alpha}'=R_{\alpha} \quad R_{\beta}'=R_{\beta} \quad
\varphi=\varphi'+\kappa\frac{1-L}{2} \quad
\chi=\chi'+\kappa(L-1) \mbox{mod}\, 2  .
\end{equation}
Remember that in the case of length independent boundary terms, the
value of $\chi$ contains the full information about the partition
function.
Observe now, that it is possible to choose values of $\kappa$ such
that the value of $\chi$ in \eref{prim} becomes independent of $L$ as long
as one considers only certain sequences of lattice lengths.
This is  exactly the case
if $\kappa/2$ is rational, i.e. $\kappa/2=m/n, m  \in \mathbb{Z}, n\in \mathbb{N}^+$.
In  this  case $\chi$ is independent  of $L$ for $L=ln+r, 0\leq r < n, l\in \mathbb{N}$,
i.e.
\begin{equation}
\label{chip}
\chi=\chi'+  \frac{2  m(r-1)}{n}.
\end{equation}
The results obtained for the symmetric case can be adopted immediately.
This is  not  possible for  irrational values of  $\kappa/2$.
Note that this kind of commensurability and incommensurability has also
 been observed for the
periodic chain with this type of interaction \cite{AlcWre}.
\subsection{$p,q \in \mathbb{R}$}
Without loss of generality, we are going to restrict ourselves to $p>q$ which
implies $Q<1$.
We have to distinguish two different situations.
If $\alpha_-'\beta_+'$ equals zero we can adopt the results
for the symmetric case with $F=C=0$ (see appendix A for details).
One has just to exchange $\alpha_{\pm}$ and $\beta_{\pm}$ by 
$\alpha_{\pm}'$ and $\beta_{\pm}'$.
If otherwise
$\alpha_-'\beta_+'\neq 0$ the situation changes.
The energy gaps suggest the partition function for the long
chain \eref{Hlong} with boundary terms given by \eref{boundaryDM}, i.e.
\begin{eqnarray}
{\cal Z}_{\rm long}=\tr z^{ \frac{2L}{(Q+1/Q)\pi}(H - e_{\infty}L-f_{\infty}) }
=\frac{2}{\eta(z)}
\sum_{m \in \mathbb{Z}/2} z^{2(m+\frac{\Delta_x}{2})^2+ 2mL\Delta_y}
\label{puq}
\end{eqnarray}
where
\begin{equation}
\label{Deltax}
\Delta_x=\frac{\rmi}{2\pi}\ln\Gamma \qquad \Delta_y= \frac{\rmi}{\pi}\frac{Q-Q^{-1}}{Q+Q^{-1}}
\end{equation}
and
\begin{eqnarray}
\label{Gamma}
\Gamma&=
&(1-Q^2(1-2\alpha_z^2-2\alpha_-'\alpha_+')-2\alpha_z^2Q^4) \nonumber \\
&&\times (1-Q^2(1-2\beta_z^2-2\beta_-'\beta_+')-2\beta_z^2Q^4)/(\alpha_-'^2\beta_+'^2Q^4(1+Q^2)^2) .
\end{eqnarray}
Note that the value of $\Delta_y$ is purely imaginary.
Hence the length dependent term given in the partition function is a phase.
Such a term already appeared in the toroidal partition function
for the asymmetric model with
periodic boundary conditions \cite{NohKim}. 
Note also that the result \eref{puq}
 simplifies to  the  partition function we obtained in the Neumann-Neumann case for $H_{\rm long}$
 if one sets  $Q=1$ (see \eref{zlongnn}).

In \cite{paper1} we computed the exact ground-state energy on the finite chain for boundaries defined by
\begin{equation}
\label{dddd}
\alpha_z=\beta_z=0 \qquad \alpha_+'\alpha_-'=\beta_+'\beta_-'=1.
\end{equation}
If  we  introduce the  parameter
$\chi'$  via $\beta_+'=\rme^{\rmi\pi\chi'} \alpha_+'$,
then  
\eref{Gamma} simplifies to
\begin{equation}
\Gamma=\left(\rme^{\rmi\pi\chi'}Q^2\right)^{-2} .
\end{equation}
Expanding the exact expression obtained in \cite{paper1} leads to
\begin{eqnarray}
\label{puqe0}
\fl
E_0\sim -\frac{Q+Q^{-1}}{2\pi}L - \frac{Q+Q^{-1}+(Q-Q^{-1})\frac{1}{2}\ln\Gamma}{2\pi}
     -  \frac{(Q+Q^{-1})\pi}{2 L}\left(\frac{1}{24} +\frac{(\ln  \Gamma)^2}{8\pi}\right)  .
\end{eqnarray}
It is  only for this case that we can perform the projection 
onto the $(+,+)$-sector.
 We obtain 
\begin{equation}
\label{puqodd}
{\cal Z}=\frac{1}{\eta(z)}
\sum_{m \in \mathbb{Z}}z^{2(m+\chi'/2-\rmi \ln Q/\pi)^2+2mL\Delta_y}
\quad \mbox{for odd $L$}
\end{equation}
\begin{equation}
\label{puqeven}
{\cal Z}=\frac{1}{\eta(z)}
\sum_{m \in \frac{2\mathbb{Z}+1}{2}}z^{2(m+\chi'/2-\rmi \ln Q/\pi)^2+2mL\Delta_y}
\quad \mbox{for even $L$}.
\end{equation}
Note that our results for $H_{\rm long}$ only apply if \eref{Gamma} is  different from zero.
If the  nominator  of $\Gamma$ vanishes we obtain logarithmic terms in the asymptotic 
 behaviour  of the energy gaps
similar the ones obtained for the symmetric chain.
We obtained the fermion energies
\begin{eqnarray}
\fl
2\Lambda  \sim \frac{Q+Q^{-1}}{2L} \Biggl\{ k\pi \pm \frac{1}{2}\Biggl[ \arg \Delta 
-\rmi\ln L 
\Biggr]\Biggr\}\pm \rmi \frac{Q-Q^{-1}}{2} 
\end{eqnarray}
where
\begin{equation}
\Delta=\frac{2\alpha_-'^2\beta_+'^2 Q^4 (1+Q^2)^2}{2AQ^2+4Q^4(B+E^2)-6Q^6(D+2E^2)+8E^2Q^8}.
\label{deltapuq}
\end{equation}
The values of $A,B,D,E$ are given in \eref{koeff}.
We  are not  going to consider the case, where the denominator in \eref{deltapuq}
vanishes.

\section{Conclusions}
In this paper we considered the spectra of the XX--model  with boundary fields 
given by the Hamiltonian $H$ in \eref{HXX}. 
In order to obtain our results we also studied the Hamiltonian  $H_{\rm long}$ 
defined by \eref{Hlong}. Furthermore we considered the asymmetric XX--chain \eref{Ha}.
Here we separated two cases. First we considered the case where the values of $p$ and $q$
in \eref{Ha} are given by $p=\rme^{\rmi\pi \kappa}/2$ and $q=\rme^{-\rmi \pi \kappa}/2$
(this corresponds to Dzyaloshinsky-Moriya interactions). Second we considered 
the case where $p$ and $q$ are real numbers.

For periodic boundary conditions the partition function for $H$ is given by the
partition function of the free boson with periodic boundary conditions \eref{bosonp}
\cite{AlcBaaGriRit2,AlcBarBat,AlcBarBat2}.
The partition function for $H_{\rm long}$ with periodic boundary conditions 
is just the expression \eref{bosonp} multiplied by 4. The spectra of the asymmetric chain
with Dzyaloshinsky-Moriya interactions and periodic boundary conditions have been studied
in \cite{AlcWre}, whereas the partition function for
 real values of $p$ and $q$ and periodic boundary conditions 
was given in \cite{NohKim}.

The results we obtained for $H$ and $H_{\rm long}$ can be encoded in terms of the parameters
given in \eref{newparam}. For reasons discussed in the text we obtained the energy gaps
only for the cases given in \tref{structure}.
Note that as long as we restrict ourselves to hermitian boundary terms
we studied the most general case. This is not true for non-hermitian boundaries.

For $H$ with hermitian boundaries we obtained the partition functions corresponding 
to one of the three boson partition functions \eref{bosondd},\eref{bosonnn} and
\eref{bosondn}. 
However, we found these partition functions also for certain non-hermitian boundary terms.

We obtained the Neumann-Neumann partition function \eref{bosonnn}
if all non-diagonal terms
are present, where the value of $\Delta$ in \eref{bosonnn} is given by \eref{deltann}.
The Dirichlet-Neumann
partition function \eref{bosondn} is obtained for boundary terms which satisfy \eref{dnnonhermparam}.
We found the Dirichlet-Dirichlet partition function \eref{bosondd} for two types of boundary terms 
given by \eref{dd2} respectively \eref{ddnonherm}.
The value of $\Delta$ in \eref{bosondd} is given by \eref{deltadd} in both cases.
Furthermore, for the case of non-hermitian boundaries we found a case which is special.
For boundary terms given by \eref{pa} the partition function is the Dirichlet-Dirichlet partition
function \eref{bosondd} multiplied by 2, where the value of $\Delta$ is $0$ or $1/2$ for
odd respectively even values of the lattice length.

The partition functions for $H_{\rm long}$ have also been considered. If all non-diagonal
terms are present it is given by \eref{zlongnn}. For the other cases it is just the
partition function of $H$ multiplied by 4. We also computed the values of the free surface 
energies of $H_{\rm long}$ and $H$ for certain boundary terms. They are given in \tref{tabfac}.
In this table we have also 
given the values of the lowest highest weights which appear in the spectra of $H_{\rm long}$ for
these boundary terms.

For the last two cases in \tref{structure} we obtained logarithmic corrections to
the free surface energy (equation \eref{loge0}). Here we found also logarithmic terms 
in the asymptotic behaviour of the energy gaps (equation \eref{aslog}).

In the case of Dzyaloshinsky-Moriya interactions, we restricted ourselves 
to hermitian chains. The value of the phase $\kappa$ (see \eref{kk}) has only an
effect on the spectra if non-diagonal boundary terms are present at both ends of the chain.
In this case we obtain the Neumann-Neumann partition function \eref{bosonnn} if
$\kappa/2$ is a rational number.
The value of $\chi$ in the definition of $\Delta$ in \eref{deltann} has just to be 
exchanged by the expression in \eref{chip}.

The partition function for the asymmetric chain \eref{Ha} with real values of $p$ and $q$
has only been obtained for one special type of boundaries (see equation \eref{dddd}).
It is given by \eref{puqodd} for $L$ odd respectively by \eref{puqeven} for $L$ even.
The asymptotic behaviour 
of the ground-state energy for this case is given in \eref{puqe0}.

\ack  
I would like to thank Birgit Wehefritz for her contribution to the early stage of this work.
I would also like to thank Vladimir Rittenberg for many discussions and constant encouragement.
I would like to thank Paul Pearce for helpful comments.
I am grateful to Klaus Krebs for carefully reading the manuscript and many fruitful 
discussions. This work was supported by the TMR Network Contract FMRX-CT96-0012 of the European 
Commission.  
\appendix
\section{Determination of the energy gaps}
In \cite{paper1} we have seen that the spectrum of
$H_{\rm long}$ is given in terms of free fermions.
The fermionic energies $2\Lambda_n$ are given by
\begin{equation}
2\Lambda_n=\frac{1}{2}(x_n+x_n^{-1}).
\label{egap}
\end{equation}
where the $x_n$ are the roots of the polynomial
\begin{eqnarray}
\label{pol}
\fl
p(x^2)  =  \Bigl[ x^{4L+8}+1 - A (x^{4L+6}+x^2)
+(B+E^2)(x^{4L+4} +x^4)  \nonumber \\
\lo
+ (D+2E^2) (x^{4L+2}+x^6) +E^2 (x^{4L} +x^8) -2E  (x^{2L+8}+x^{2L})
\nonumber\\ \lo{+}
\left((A-B-D-1)/2-(-1)^{L} C - 2 E^2\right) (x^{2L+6}+x^{2L+2})
\nonumber \\
 \lo  + \left(A-B-D-1+ 2 (-1)^L C+ 4 E -4 E^2\right) x^{2L+4}  \left. \Bigr]
\right/(x^2-1)^2
\end{eqnarray}
where the coefficients appearing in \eref{pol} are functions of the boundary parameters:
\begin{eqnarray}
\label{koeff}
A  =  2 (\alpha_-\alpha_+ + \beta_-\beta_+ + \alpha_z^2 + \beta_z^2-1)\qquad
C  =   \alpha_-^2 \beta_+^2+ \alpha_+^2 \beta_-^2 \nonumber \\
B  =  ( 2 \alpha_-\alpha_+-1)( 2 \beta_-\beta_+-1) +
 4 \beta_z^2(\alpha_-\alpha_+-1) +4\alpha_z^2 ( \beta_+\beta_- -1)
\nonumber\\
 D  =  \beta_z^2 (4 \alpha_-\alpha_+ -2)  +  \alpha_z^2( 4 \beta_+\beta_- -2)  \qquad
 E  =  2 \alpha_z \beta_z .
\end{eqnarray}
Since there appear
4 zeros for each fermion, namely $x_n,x_n^{-1},-x_n$ and $-x_n^{-1}$,
the polynomial yields $L+1$ fermionic energies (see \cite{paper1} for details).
In addition to these fermions there always exists a fermion with energy $2\Lambda_0=0$,
which we named 'spurious' zero mode.
All possible combinations of these $L+2$ fermions build up the spectrum
of $H_{\rm long}$. The spectrum of $H$ is then obtained by excluding the 'spurious'
zero mode from the set of fermion energies and then taking the sector with an even or with an odd number
of fermions being excited with respect to the vacuum. How to decide 
whether one has to pick up the even or the odd sector has been discussed in detail in \cite{paper1}.

Since in this paper we are interested into the large $L$ behaviour
of the low lying energy levels, we look for the
asymptotics of the zeros of $p(x^2)$ in the vicinity of the point $x=\pm \rmi$
using the ansatz $x=\rme^{\rmi \frac{\pi}{2}-\rmi\frac{\phi}{L}}$,
where we assume $\phi$ to be a constant in leading order.
For the first four cases given in \tref{structure} we obtained the following 
equations:
\begin{itemize}
\item[$\circ$]
$F\neq 0$
\begin{equation}
\cos(2\phi)+\frac{2C}{F}=0
\end{equation}
\item[$\circ$]
$F=C=0,G\neq 0$
\begin{equation}
\phi\sin(2\phi)=0
\end{equation}
\item[$\circ$]
$F=C=G=0,K\neq 0$
\begin{equation}
\phi^2\left[\cos(2\phi)+(-1)^{L+1}\frac{J}{K}\right] =0
\end{equation}
\item[$\circ$]
$F=C=G=K=J=0$
\begin{equation}
\phi^4\exp(2\rmi\phi)=(-1)^L
\end{equation}
\end{itemize}
Solving these equations for $\phi$ yields the energy gaps (cf. \eref{egap}) in leading order.
We obtained the following expressions:
\begin{itemize}
\item[$\circ$]
$F\neq 0$
\begin{equation}
2\Lambda\sim\frac{1}{L}\left[ \frac{2n-1}{2}\pi \pm \frac{1}{2}\mbox{arccos}
\left(\frac{2C}{F}\right) \right] \qquad 1\leq n
\end{equation}
\item[$\circ$]
$F=C=0,G\neq 0$
\begin{equation}
2\Lambda\sim \frac{1}{L}\frac{n}{2}\pi \qquad 0\leq n
\end{equation}
\item[$\circ$]
$F=C=G=0,K\neq 0$
\begin{equation}
2\Lambda \sim \frac{1}{L}\left[  \frac{2n-1}{2}\pi \pm \frac{1}{2}\mbox{arccos}
\left((-1)^{L+1}\frac{J}{K}\right) \right] \qquad 1\leq n
\end{equation}
In addition to these modes there appears an additional zero mode.
\item[$\circ$]
$F=C=G=K=J=0$
\begin{equation}
2\Lambda\sim\frac{1}{L}n\pi \qquad 1\leq n \qquad \mbox{for $L$ even}
\end{equation}
\begin{equation}
2\Lambda\sim\frac{1}{L}\frac{2n-1}{2}\pi \qquad 1\leq n \qquad  \mbox{for $L$ odd}
\end{equation}
Furthermore we found 3 additional zero modes for $L$ even and 2 additional zero 
modes for $L$ odd.
\end{itemize}
For the last two cases in \tref{structure}
the asymptotic zeros of the polynomial can only be found
using the ansatz $x=\rme^{\rmi\frac{\pi}{2}-\rmi\frac{\phi}{L}}$ if
we assume the imaginary part of $\phi$ to diverge as $L$ goes to infinity.
Hence we generalize our ansatz to $x=\rme^{\rmi \frac{\pi}{2}-\rmi\frac{\phi(L)}{L}}$,
where $\phi(L)$ is a complex function.
Furthermore we assume
$\lim_{L\to \infty}\frac{\phi(L)}{L}=0$ and $\lim_{L \to \infty} \rme^{-\rmi\phi(L)}=0$.
The second assumption will be explained shortly.
In both cases our ansatz leads to the solution of 
an equation of the form 
\begin{equation}
\label{lam}
a\rme^{-2\rmi\phi(L)}\left[1+\Or\left(\frac{\phi(L)}{L}\right)\right]+\frac{2\rmi\phi(L)}{L}
\left[b+\Or\left(\frac{\phi(L)}{L}\right)\right]=0 .
\end{equation}
From this expression one can see that our second assumption is indeed necessary to
solve this equation, since otherwise the first term would be finite 
for all values of $L$, whereas the second term vanishes as $L$ goes to infinity.
Neglecting the terms of order 
 $\phi(L)/L$ this equation is solved
in terms of the so called Lambert W function ${\cal L}$  which
is defined by the property  ${\cal L}(x) \rme^{{\cal L}(x)}=x$.
We obtain
\begin{equation}
\phi(L)=-\rmi{\cal L}(\Delta L)/2
\quad
\mbox{where}
\quad
\Delta=-\frac{a}{b} .
\end{equation}
The asymptotic behaviour of ${\cal L}$ is well known \cite{LambertW}, i.e.
\begin{equation}
{\cal L}(L)\sim 2\rmi k\pi +\ln L -\ln(\ln L+2\rmi \pi k) + ...
\end{equation}
Note that this expression is in accordance with our assumptions we made concerning
the asymptotic behaviour of $\phi(L)$.

In this paper we also considered the asymmetric chain, which is similar
to the symmetric chain with length dependent boundary parameters (cf. \eref{boundaryDM}).
This length dependence enters the polynomial \eref{pol} only
via the coefficient $C$, which becomes
\begin{equation}
C= \alpha_-'^2\beta_+'^2 Q^{2-2L}+\alpha_+'^2\beta_-'^2 Q^{2L-2} .
\label{CQ}
\end{equation}
For $Q<1$ (the case considered in section 3.2) the first term 
 on the RHS of \eref{CQ} diverges exponentially whereas
the second term vanishes exponentially as a function of the lattice length $L$.
Hence we have to distinguish two different situations.
If $\alpha_-'\beta_+'$ equals zero we may find the asymptotic zeros of the polynomial
as for the symmetric case with $F=C=0$.
If otherwise
$\alpha_-'\beta_+'\neq 0$ then $C$ diverges as $L$  is increased.
This has to be compensated by modifying our ansatz to 
$x=\rme^{\rmi \frac{\pi}{2}-\rmi\frac{\phi}{L}+\ln Q}$ .
This works as long as the nominator of $\Gamma$ in \eref{Gamma} is 
different from zero.
Otherwise we have to modify our ansatz another time to
$x=\rme^{\rmi \frac{\pi}{2}-\rmi\frac{\phi(L)}{L}+\ln Q}$, where
$\lim_{L\to \infty}\frac{\phi(L)}{L}=0$ and $\lim_{L \to \infty} \rme^{-\rmi\phi(L)}=0$.
This leads again to an equation of the form \eref{lam} which can be solved as described above.

\end{document}